\documentclass[3p,times,fleqn]{elsarticle}
 \biboptions{comma,sort&compress}
 
\usepackage{graphicx}
\usepackage{amsmath}
\usepackage{here}
\usepackage{ecrc}

\usepackage{slashed}
\usepackage{bm}
\usepackage{float}

\volume{00}

\firstpage{1}

\journalname{Nuclear and Particle Physics Proceedings}

\runauth{Marco Frasca}

\jid{nppp}

\jnltitlelogo{Nuclear and Particle Physics Proceedings}

\usepackage{amssymb}
\usepackage[figuresright]{rotating}

\begin{document}

\begin{frontmatter}

%%
%%%%%%%%%%%%%%%%%%%%%%%%%%%%%%%%%%%%%%%%%%%%%%%%%
\title{ Quark confinement in QCD in the 't Hooft limit }
 % \corref{cor0}}
 \cortext[cor0]{Talk given at 25th International Conference in Quantum Chromodynamics (QCD 22),  4 - 7 July 2022, Montpellier - FR}
 \author[label1]{Marco Frasca\fnref{fn1}}
 \fntext[fn1]{Speaker, Corresponding author.}
%  \cortext[cor0]{FAPESP CNPq-Brasil PhD student fellow.}
\ead{marcofrasca@mclink.it}
\address[label1]{Via Erasmo Gattamelata, 3, 
00176 Rome (Italy)}
\author[label2]{Anish Ghoshal}
\address[label2]{INFN, Rome, Italy and Warsaw University, Poland}
\author[label3]{Stefan Groote}
\address[label3]{University of Tartu, Estonia}
%\address[label2]{Laboratoire
%Particules et Univers de Montpellier, CNRS-IN2P3, 
%Case 070, Place Eug\`
%Bataillon, 34095 - Montpellier, France.}
% \author[label3]{F. Fanomezana\corref{cor1}}
%  \cortext[cor1]{PhD student.}
%\ead{fanfenos@yahoo.fr}
%\address[label3]{Institute of High-Energy Physics of Madagascar (iHEP-MAD), University of Antananarivo, 
%Madagascar}
% \author[label2,label4]{S. Narison\fnref{fn1}}
%   \fntext[fn1]{Speaker, Corresponding author.}
%    \ead{snarison@yahoo.fr}

%\address[label4]{Madagascar consultant of the Abdus Salam International Centre for Theoretical Physics (ICTP), via Beirut 6,34014 Trieste, Italy .}
% \author[label3]{A. Rabemananjara\corref{cor1}}
%  \cortext[cor2]{Ph.D. student}
%\ead{achris\_01@yahoo.fr}

\pagestyle{myheadings}
\markright{ }
\begin{abstract}
We treat quantum chromodynamics (QCD) using a set of Dyson-Schwinger equations derived, in differential form, with the Bender-Milton-Savage technique. In this way, we are able to derive the low energy limit that assumes the form of a non-local Nambu-Jona-Lasinio model. The corresponding gap equation is then studied to show that such a model has no free quarks in the low-energy limit.
\end{abstract}
% \begin{document}
\begin{keyword}  
%% keywords here, in the form: keyword \sep keyword

%% MSC codes here, in the form: \MSC code \sep code
%% or \MSC[2008] code \sep code (2000 is the default)

\end{keyword}

\end{frontmatter}
%%%%%%%%%%%%
%\vspace*{-1.5cm}
\section{Introduction}

Understanding quark confinement is one of the most outstanding problem in QCD. Some criteria have been devised (e.g. \cite{Kugo:1977zq,Kugo:1979gm}) but a first principle proof is not known yet. Some theory are shown to confine as in supersymmetric Yang-Mills theory \cite{Novikov:1983uc,Shifman:1986zi,Ryttov:2007cx} or standard Yang-Mills theory \cite{Chaichian:2018cyv} where the exact beta function was obtained. Indeed, the gluon propagator is also known in closed form with more or less fitting parameters \cite{Cornwall:1981zr,Cornwall:2010bk,Dudal:2008sp,Frasca:2015yva,Frasca:2017slg}). It should be emphasized that it is essential to obtain the low-energy limit of QCD from first principles as this opens up a wealthy number of applications in several fields ranging from nuclear physics to cosmology. With the given results in Yang-Mills theory, this can be accomplished. The relevant approximations involved are strong coupling limit and 't Hooft limit of number of colors running to infinity keeping the product of the number of colors and the square of coupling constant \cite{tHooft:1973alw,tHooft:1974pnl}. We will obtain such a limit and prove quark confinement in the 't Hooft limit \cite{Frasca:2022lwp}.

\section{Bender-Milton-Savage technique}

Our approach is based on the Bender-Milton-Savage (BMS) technique that permits to derive the Dyson-Schwinger equations in PDE form \cite{Bender:1999ek}. This technique can be better explained referring to a scalar field. Therefore, we consider the following partition function
\begin{equation}
    Z[j]=\int[D\phi]e^{iS(\phi)+i\int d^4xj(x)\phi(x)}.
\end{equation}
We start from the equation of motion for th 1P-function
\begin{equation}
\left\langle\frac{\delta S}{\delta\phi(x)}\right\rangle=j(x),
\end{equation}
assuming
\begin{equation}
\left\langle\ldots\right\rangle=\frac{\int[D\phi]\ldots e^{iS(\phi)+i\int d^4xj(x)\phi(x)}}{\int[D\phi]e^{iS(\phi)+i\int d^4xj(x)\phi(x)}}
\end{equation}
Then, we set $j=0$ to obtain the equation for the 1P-function. We derive this equation with respect to $j$ to obtain the equation for the 2P-function.The definition for the nP-functions is the following
\begin{equation}
\langle\phi(x_1)\phi(x_2)\ldots\phi(x_n)\rangle=\frac{\delta^n\ln(Z[j])}{\delta j(x_1)\delta j(x_2)\ldots\delta j(x_n)}.
\end{equation}
This implies
\begin{equation}
\frac{\delta G_k(\ldots)}{\delta j(x)}=G_{k+1}(\ldots,x).
\end{equation}

This procedure can be iterated to any desired order. These equations have the great advantage that permit to use possible exact solutions to them providing closed form formulas for the correlation functions of the theory.

\section{1P and 2P functions for QCD}

In order to make our computations simpler, as done in Ref.~\cite{Frasca:2015yva}, we evaluate our equations in the Landau gauge.

The Bender-Milton-Savage method yields for the 1P-functions
\begin{eqnarray}
      &&\partial^2G_{1\nu}^{a}(x)+gf^{abc}(
		\partial^\mu G_{2\mu\nu}^{bc}(0)+ \nonumber \\
		&&\partial^\mu G_{1\mu}^{b}(x)G_{1\nu}^{c}(x)-
		\partial_\nu G_{2\mu}^{\nu bc}(0)
		\nonumber \\
		&&-\partial_\nu G_{1\mu}^{b}(x)G_{1}^{\mu c}(x)) \nonumber \\
		&&+gf^{abc}\partial^\mu G_{2\mu\nu}^{bc}(0)+gf^{abc}\partial^\mu(G_{1\mu}^{b}(x)G_{1\nu}^{c}(x))
		\nonumber \\
		&&+g^2f^{abc}f^{cde}(G_{3\mu\nu}^{\mu bde}(0,0)
		+G_{2\mu\nu}^{bd}(0)G_{1}^{\mu e}(x)
		\nonumber \\
	    &&+G_{2\nu\rho}^{eb}(0)G_{1}^{\rho d}(x)
	    +G_{2\mu\nu}^{de}(0)G_{1}^{\mu b}(x)+ \nonumber \\
	    &&G_{1}^{\mu b}(x)G_{1\mu}^{d}(x)G_{1\nu}^{e}(x)) \nonumber \\
		&&=g\sum_{q,i}\gamma_\nu T^aS_{q}^{ii}(0)+g\sum_{q,i}{\bar q}_1^i(x)\gamma_\nu T^a q_1^i(x),
\end{eqnarray}
and for the quarks
\begin{equation}
	(i\slashed\partial-m_q)q_{1}^{i}(x)+g{\bm T}\cdot\slashed{\bm G}_1(x) q_{1}^{i}(x) 
	+g{\bm T}\cdot\slashed{\bm W}^{i}_q(x,x)= 0.
\end{equation}

Here and in the following Greek indexes ($\mu,\nu,\ldots$) are for the space-time and Latin index ($a, b,\ldots$) for the gauge group. We recognize immediately a known property of the Dyson-Schwinger equations that equations for the lower order correlation functions depend on values of higher order correlation functions.

At this stage, we assume the mapping theorem as done in Ref.~\cite{Frasca:2015yva}. So, we assume 
\begin{equation}
G_{1\nu}^a(x)\rightarrow\eta_\nu^a\phi(x)
\end{equation}
being $\phi(x)$ a scalar field. Let us introduce the $\eta$-symbols as follows
\begin{eqnarray}
\eta_\mu^a\eta^{a\mu} &=& N^2-1. \nonumber \\ 
\eta_\mu^a\eta^{b\mu} &=& \delta_{ab}, \nonumber \\
\eta_\mu^a\eta_\nu^a &=& \left(g_{\mu\nu}-\delta_{\mu\nu}\right)/2.
\end{eqnarray}
These simplify the equations down to
\begin{eqnarray}
&&\partial^2\phi(x)+2Ng^2\Delta(0)\phi(x)+Ng^2\phi^3(x) 
\nonumber \\
&&=\frac{1}{N^2-1}\left[g\sum_{q,i}\eta^{a\nu}\gamma_\nu T^aS_{q}^{ii}(0)\right. \nonumber \\
&&\left.+g\sum_{q,i}{\bar q}_1^i(x)\eta^{a\nu}\gamma_\nu T^a q_1^i(x)\right]
\nonumber \\
&&(i\slashed\partial-m_q^i)q_{1}^{i}(x)+g{\bm T}\cdot\slashed\eta\phi(x) q_{1}^{i}(x) = 0.
\end{eqnarray}

We do the same for the 2P-functions. In the Landau gauge, for the gluon 2P-function we get 
\begin{equation}
    G_{2\mu\nu}^{ab}(x-y)=\left(\eta_{\mu\nu}-\frac{\partial_\mu\partial_\nu}{\partial^2}\right)\Delta_\phi(x-y)
\end{equation}
being $\eta_{\mu\nu}$ the Minkowski metric, and $\Delta_\phi(x-y)$ is the propagator of the $\phi$ given the map between the scalar and the Yang-Mills fields. Finally, we can write
\begin{eqnarray}
&&\partial^2\Delta_\phi(x-y)+2Ng^2\Delta_\phi(0)\Delta_\phi(x-y)+3Ng^2\phi^2(x)\Delta_\phi(x-y) \nonumber \\
&&=g\sum_{q,i}{\bar Q}^{ia}_\nu(x-y)\gamma^\nu T^a q_{1}^{i}(x)
\nonumber \\
&&+g\sum_{q,i}{\bar q}_1^{i}(x)\gamma^\nu T^a Q^{ia}_\nu(x-y) + \delta^4(x-y)\nonumber \\
&&\partial^2 P^{ad}_2(x-y)=\delta_{ad}\delta^4(x-y) \nonumber \\
&&(i\slashed\partial-m_q^i)S^{ij}_q(x-y) \nonumber \\
&&+g{\bm T}\cdot\slashed\eta\phi(x) S^{ij}_q(x-y)=\delta_{ij}\delta^4(x-y) \nonumber \\  
&&\partial^2W_{q\nu}^{ai}(x-y)+2Ng^2\Delta_\phi(0)W_{q\nu}^{ai}(x-y)+3Ng^2\phi^2(x)W_{q\nu}^{ai} \nonumber \\
&&=g\sum_{j}{\bar q}_1^{j}(x)\gamma_\nu T^a S^{ji}_q(x-y)\nonumber \\
&&(i\slashed\partial-m_q^i)Q^{ia}_\mu(x-y)+g{\bm T}\cdot\slashed\eta\phi(x) Q^{ia}_\mu(x-y) \nonumber \\
&&+gT^a\gamma_\mu\Delta_\phi(x-y) q_{1}^{i}(x)=0.
\end{eqnarray}

\section{'t Hooft limit}

't Hooft limit means to solve the theory assuming \cite{tHooft:1973alw, tHooft:1974pnl}
\begin{equation}
  N\rightarrow\infty,\qquad Ng^2=constant, \qquad Ng^2\gg 1.
\end{equation}
The gauge group is SU(N) being $N$ is the number of colors. To solve the equations in the strong coupling limit, we need a proper perturbation technique. We proposed such a method in Ref.\cite{Frasca:2013tma}. We rescale $x\rightarrow\sqrt{Ng^2}x$ and write the equation for the gluon field as follows
\begin{eqnarray}
      \partial^2\phi(x')+2\Delta_\phi(0)\phi(x')+3\phi^3(x')&=& \\
\frac{1}{\sqrt{Ng^2}\sqrt{N}(N^2-1)}\left[\sum_{q,i}\eta\cdot\gamma\cdot TS_{q}^{ii}(0)+\right.&& \nonumber \\
\left.\sum_{q,i}{\bar q}_1^i(x')\eta\cdot\gamma\cdot T q_1^i(x')\right].&&\nonumber
\end{eqnarray}
In the 't Hooft limit, the equation for the 1P-function becomes
\begin{eqnarray}
	\partial^2\phi_0(x)+2Ng^2\Delta_\phi(0)\phi_0(x)+3Ng^2\phi_0^3(x)=0,& \nonumber \\
		(i\slashed\partial-m_q^i){\hat q}_{1}^{i}(x)+g{\bm T}\cdot\slashed{\eta}\phi(x) q_{1}^{i}(x)=0.&
\end{eqnarray}
At the leading order the only effect is seen on masses. Therefore, we can solve the equation for the gluon field taking
\begin{eqnarray}
\phi_0(x)=\sqrt{\frac{2\mu^4}{m^2+\sqrt{m^4+2Ng^2\mu^4}}}\times
\nonumber \\
{\rm sn}\left(p\cdot x+\chi,\kappa\right),
\end{eqnarray}
being sn a Jacobi elliptical function, $\mu$ and $\chi$ arbitrary integration constants and $m^2=2Ng^2\Delta_\phi(0)$ a mass shift arising from quantum corrections. We have
\begin{equation}
\kappa=\frac{-m^2+\sqrt{m^4+2Ng^2\mu^4}}{-m^2-\sqrt{m^4+2Ng^2\mu^4}}.
\end{equation}
This is true provided that the following dispersion relation holds
\begin{equation}
    p^2=m^2+\frac{Ng^2\mu^4}{m^2+\sqrt{m^4+2Ng^2\mu^4}}.
\end{equation}
For the equations of the 2P-functions one has
\begin{eqnarray}
\partial^2\Delta_\phi(x,y)+2Ng^2\Delta_\phi(0)\Delta(x-y)+3Ng^2\phi_0^2(x)\Delta_\phi(x-y) \nonumber \\
=g\sum_{q,i}{\bar Q}^{ia}_\nu(x,y)\gamma^\nu T^a {\hat q}_{1}^{i}(x)
\nonumber \\
+g\sum_{q,i}{\bar{\hat q}}_1^{i}(x)\gamma^\nu T^a Q^{ia}_\nu(x,y)
+ \delta^4(x-y) \nonumber \\
\partial^2 P^{ad}_2(x-y)=\delta_{ad}\delta^4(x-y) \nonumber \\
(i\slashed\partial-m_q^i){\hat S}^{ij}_q(x-y)+g{\bm T}\cdot\slashed\eta\phi(x) S^{ij}_q(x-y)=\delta_{ij}\delta^4(x-y) \nonumber \\  
\partial^2W_{q\nu}^{ai}(x,y)+2Ng^2\Delta_\phi(0)W_{q\nu}^{ai}(x,y)+3Ng^2\phi_0^2(x)W_{q\nu}^{ai}(x,y)\nonumber \\
=g\sum_{j}{\bar {\hat q}}_1^{j}(x)\gamma_\nu T^a {\hat S}^{ji}(x-y) \nonumber \\
(i\slashed\partial-{\hat M}_q^i){\hat Q}^{ia}_\mu(x,y)+gT^a\gamma_\mu\Delta_\phi(x-y) {\hat q}_{1}^{i}(x)=0.
\end{eqnarray}
These equations can be solved by finding a solution to the following equation
\begin{eqnarray}
\partial^2\Delta_0(x-y)+[m^2+3Ng^2\phi_0^2(x)]\Delta_0(x-y)&=&
\nonumber \\
\delta^4(x-y).&&
\end{eqnarray}
In momentum space, the solution of this equation is given by \cite{Frasca:2015yva,Frasca:2013tma}
\begin{eqnarray}
   \Delta_0(p)=M{\hat Z}(\mu,m,Ng^2)\frac{2\pi^3}{K^3(\kappa)}\times \nonumber \\
	\sum_{n=0}^\infty(-1)^n\frac{e^{-(n+\frac{1}{2})\pi\frac{K'(\kappa)}{K(\kappa)}}}
	{1-e^{-(2n+1)\frac{K'(\kappa)}{K(\kappa)}\pi}}\times \nonumber \\
	(2n+1)^2\frac{1}{p^2-m_n^2+i\epsilon}
\end{eqnarray}
being
\begin{equation}
M=\sqrt{m^2+\frac{Ng^2\mu^4}{m^2+\sqrt{m^4+2Ng^2\mu^4}}},
\end{equation}
and ${\hat Z}(\mu,m,Ng^2)$ a given constant. The spectrum is given by $m_n$ and a proper gap equation \cite{Frasca:2017slg}.

\section{Non-local Nambu-Jona-Lasinio approximation}

\begin{figure}[H]
 \includegraphics[width=.4\textwidth]{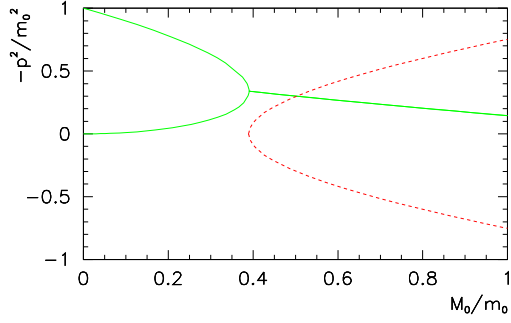}
 \label{fig1}
 \caption{$M_0$ is the effective quark mass, $m_0$ is the mass of the lowest glue state.}
\end{figure}

% Let us introduce the self-energy
% \[
% \blu{
% \Sigma(x,x)=g^2\int d^4y'\Delta_0(x-y')T^a\gamma^\nu\sum_{k}{\bar {\hat q}}_1^{k}(y')\gamma_\nu T^a {\hat S}^{ki}_q(y'-x)}
% \]
In the strong coupling approximation and 't Hooft limit, one gets the Dyson-Schwinger equations for the following nonlocal-NJL-model \cite{Frasca:2021zyn}:
\begin{eqnarray*}
\lefteqn{{\cal L}'_{\rm NJL}\ =\ \sum_i\bar\psi_i(x)
  (i\gamma^\mu\partial_\mu-m_q)\psi_i(x)\strut}\nonumber\\&&\strut
  +\frac{N_cg^2}2\int d^4y\Delta_0(x-y)\sum_{i,j}\bar\psi_i(x)\psi_j(y)
  \bar\psi_j(y)\psi_i(x)\strut\nonumber\\&&\strut
  +\frac{N_cg^2}2\int d^4y\Delta_0(x-y)\sum_{i,j}\bar\psi_i(x)i\gamma_5\psi_j(y)
  \bar\psi_j(y)i\gamma_5\psi_i(x)\strut\nonumber\\&&\strut
  -\frac{N_cg^2}4\int d^4y\Delta_0(x-y)\sum_{i,j}\bar\psi_i(x)\gamma^\mu
  \psi_j(y)\bar\psi_j(y)\gamma_\mu \psi_i(x)\strut\nonumber\\&&\strut
  -\frac{N_cg^2}4\int d^4y\Delta_0(x-y)\sum_{i,j}
  \bar\psi_i(x)\gamma^\mu\gamma_5\psi_j(y)
  \bar\psi_j(y)\gamma_\mu\gamma_5\psi_i(x).
\end{eqnarray*}
This model yields the following gap equation for the quark masses
\begin{equation}
M_q(p)=m_q+4N_fN_cg^2\frac{\tilde\Delta_0(p)}{\tilde\Delta_0(0)}
\int\frac{d^4p'}{(2\pi)^4}\frac{\tilde\Delta_0(p')M_q(p')}{p^{\prime2}
  -M_q^2(p')}.
\end{equation}
In order for a quark to be free, this mass should represent a pole on the real axis in the quark propagator. We will show that this is not generally true and quarks are confined. This means that the gap equation has not always a solution at decreasing energy and we move from a chiral condensate of quarks to an instanton liquid of glue excitations where quark bound states are the particles in the spectrum of the theory. The solution of the gap equation is shown graphically in Fig.~\ref{fig1}
 %\caption{$\sqrt{\sigma}=0.417\ GeV$ and $d^{-1}=0.58\ GeV$.}
In this figure the two lowest zeros of $-p^2=M_q^2(p)$ in Euclidean domain are given in units of $m_0^2$, splitted into real parts (green straight lines) and imaginary parts (red dashed lines), in dependence on the ratio $M_0/m_0$. The zeros become complex for approximately $M_0/m_0>0.39$. Given the values $\Lambda=1\ \rm{GeV}$ and $m_0$ the mass of the f0(500)\footnote{We have chosen this resonance because its possible interpretation is that it is the scalar field responsible for the chiral symmetry breaking in the NJL model.}, we are deeply in the confined regime for QCD.

\begin{figure}[H]
  \includegraphics[width=.4\textwidth]{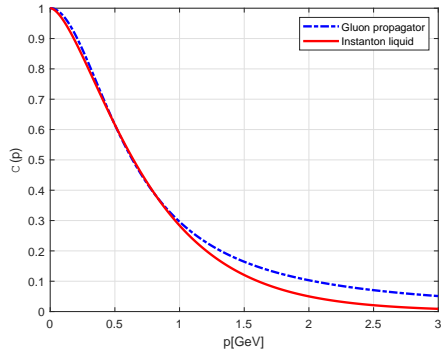}
  \caption{\it Comparison of our form factor  and that of an instanton liquid.\label{fig2}}.
\end{figure}

The evidence that the gluon propagator represents an instanton liquid quite well was given Ref.~\cite{Frasca:2013kka}. The relation between the non-local factor arising from the instanton liquid and the factor appearing in the non-local NJL model is
\begin{equation}
    \frac{2}{\tilde\Delta(0)}\tilde\Delta(p)={\cal C}(p),
\end{equation}
For an instanton liquid (normalized to zero momentum) we have (see \cite{Hell:2008cc} and refs. therein)
\begin{equation}
    {\cal C}_I(p)=p^2\left\{\pi d^2\dfrac{d}{d\xi}
  \big[I_0(\xi)K_0(\xi)-I_1(\xi)K_1(\xi)\big]\right\}^2
\end{equation}
where $\xi=\frac{|p|d}{2}$.
Indeed, the comparison yields Fig.~\ref{fig2}.
The agreement is very good and we can depict a scenario with a chiral quark condensate with massive and possibly unconfined quarks and that of an instanton liquid for the glue sector where the only particles in the spectrum are bounded quark states and colorless glue states. Free quarks are no more in the spectrum of the theory.

\section{Conclusions}

There are different approaches to understand the confinement of quarks. One of
these is given by solutions of the gap equation of the dynamical quark mass.
With a reasonable UV cutoff and fixed by the glueball
spectrum starting at the mass of the $f_0(500)$ resonance, free quarks are no more in the spectrum of QCD.
As a consequence, free quarks are no longer physical states
of the theory and the quarks can be expected to be confined in the 't~Hooft limit. 
The low-energy limit of QCD turns out
to be a well-defined non-local NJL model with all the parameters obtained from
QCD. 

Having a low-energy limit of QCD permits to do several computations to be
compared with experiments. Indeed, our first application was to the $g-2$
problem with a very satisfactory agreement with data  \cite{Frasca:2021yuu}.

\section*{Acknowledgements\label{Ack}}
The research was supported in part by the European Regional Development Fund
under Grant No.~TK133.

\end{document}